\documentclass[lettersize,journal]{IEEEtran}
\usepackage{amsmath,amsfonts}
\usepackage{algorithmic}
\usepackage{algorithm}
\usepackage{array}
\usepackage[caption=false,font=normalsize,labelfont=sf,textfont=sf]{subfig}
\usepackage{textcomp}
\usepackage{stfloats}
\usepackage{url}
\usepackage{verbatim}
\usepackage{graphicx}
\usepackage{color}
\usepackage{cite}
\begin{document}

\title{Hungarian Qubit Assignment for Optimized Mapping of Quantum Circuits on Multi-Core Architectures\vspace{-0.2cm}}

\author{Pau Escofet, Anabel Ovide, Carmen G. Almudever, Eduard Alarcón, and Sergi Abadal 
\thanks{P. Escofet, E. Alarcón, and S. Abadal are with Universitat Politècnica de Catalunya, and A. Ovide and C. G. Almudever are with Universitat Politècnica de València.}
\thanks{Authors acknowledge support from the European Research Council (ERC) under GA 101042080 (WINC) and the European Innovation Council (EIC) Pathfinder scheme, GA 101099697 (QUADRATURE).}}



\maketitle

\begin{abstract}
Modular quantum computing architectures offer a promising alternative to monolithic designs for overcoming the scaling limitations of current quantum computers. To achieve scalability beyond small prototypes, quantum architectures are expected to adopt a modular approach, featuring clusters of tightly connected quantum bits with sparser connections between these clusters. Efficiently distributing qubits across multiple processing cores is critical for improving quantum computing systems' performance and scalability. To address this challenge, we propose the Hungarian Qubit Assignment (HQA) algorithm, which leverages the Hungarian algorithm to improve qubit-to-core assignment. The HQA algorithm considers the interactions between qubits over the entire circuit, enabling fine-grained partitioning and enhanced qubit utilization. We compare the HQA algorithm with state-of-the-art alternatives through comprehensive experiments using both real-world quantum algorithms and random quantum circuits. The results demonstrate the superiority of our proposed approach, outperforming existing methods, with an average improvement of 1.28$\times$.
\end{abstract}


\section{Introduction}
\IEEEPARstart{Q}{uantum} computing has emerged as a groundbreaking paradigm with the potential to revolutionize various fields, such as cryptography, machine learning, and optimization \cite{cryptography, quant_mach_learn, overview}, thanks to the use of quantum properties such as entanglement and superposition.

Quantum computers \cite{feynman_1982_simulating} are currently constrained by a limited number of qubits. Even the most advanced quantum processors comprise only a few hundred qubits \cite{gambetta_2022_expanding}. To unleash the total computational power and achieve universal fault tolerance, scaling up quantum computers while ensuring minimal increments in error rates becomes imperative. Nowadays, quantum computers are designed as single-chip processors, known as single-core quantum processors, containing all qubits within a single chip. This monolithic architecture encounters challenges in scalability due to the undesirable rise in qubit interactions, leading to issues like crosstalk \cite{crosstalk}. To overcome these limitations, alternative approaches are sought, aiming to expand qubit counts and enhance scalability for future quantum systems.

\begin{figure}
    \centering
        \includegraphics[width=0.22\textwidth]{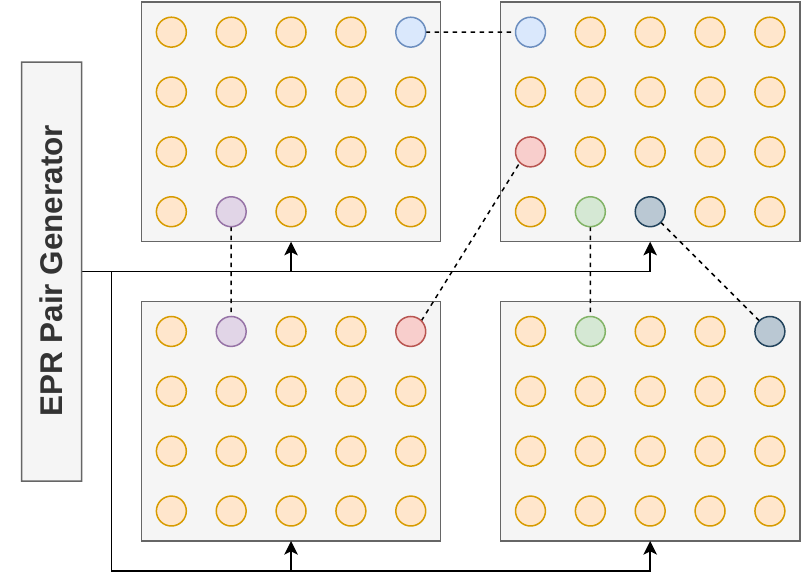}
    \hfill
        \includegraphics[width=0.22\textwidth]{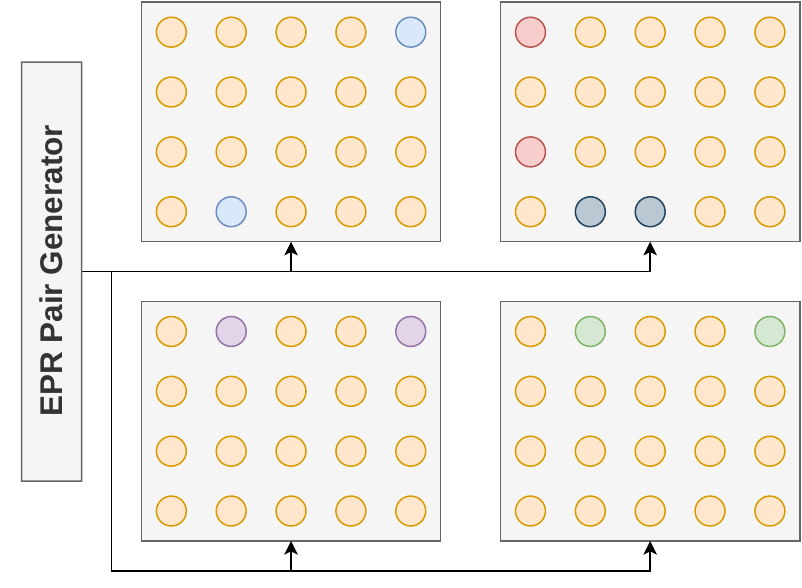}
    \vspace{-0.2cm}
    \caption{Example of a four-core (rectangles) quantum computer with 20 qubits (circles) per core. Interacting qubits are depicted with a colour different than orange and linked to the qubit they are interacting with. On the left, an invalid assignment is depicted, while on the right, a valid assignment (with all interacting pairs of qubits assigned to the same core) is shown.}
    \label{fig: assignment example}
    \vspace{-0.2cm}
\end{figure}

A feasible approach for scaling quantum computers is to go from a monolithic quantum computer to a multi-core approach \cite{laracuente_modeling, monroe_2014_largescale, gambetta_2022_expanding, smith_2022_scaling, rodrigo_2020_exploring, rodrigo_2021_on}. With the emergence of modular quantum computing architectures, new challenges arise, particularly in mapping quantum algorithms onto physical quantum devices. This compilation process is crucial for adapting a quantum algorithm to run efficiently on the desired quantum architecture, making it a central aspect of this cutting-edge technology. Figure \ref{fig: assignment example} depicts a valid (right) and an invalid (left) mapping of qubits into cores for a four-core quantum architecture.

Few multi-core oriented mapping algorithms \cite{baker_2020_timesliced, ferrari_2021_compiler, wu_2022_autocomm} have been developed to address these challenges, with strong assumptions on the cores' constraints. Their optimality in mapping quantum algorithms across multiple cores has not been rigorously examined. We propose an alternative multi-core mapping algorithm based on the observed limitations of current mapping techniques for multi-core architectures. It is designed to achieve fewer inter-core communications, facilitating improved scalability and performance. The performance of the proposed inter-core mapping algorithm is rigorously compared to existing state-of-the-art approaches.

\section{Background and Related Work}
To solve the scalability problem, several multi-core approaches have been proposed \cite{laracuente_modeling, monroe_2014_largescale, gambetta_2022_expanding, smith_2022_scaling, rodrigo_2020_exploring, rodrigo_2021_on}. The main goal is to increase the computational power of the quantum processor without increasing the crosstalk between qubits by having different quantum processors (or cores) connected via classical links (early stages) or by coherent quantum links (later stages).

In multi-core architectures, many challenges arise, particularly in executing two-qubit gates involving qubits physically situated on different quantum processors, as illustrated in Figure \ref{fig: assignment example}. It is possible to perform two-qubit gates between distant qubits in different chips under specific practical conditions \cite{gold_2021_entanglement}. However, the fidelity of inter-chip links is lower compared to intra-chip links.

Nonetheless, irrespective of the specific implementation approach, the overarching challenge lies in optimizing the number of non-local communications. This is essential for minimizing unnecessary actions, enhancing gate fidelity, and maximizing the efficiency of quantum computations. Various inter-core communications protocols to perform two-qubit gates exist, including remote gate operations and the transfer of quantum states followed by local gate executions.

In the context of our paper, we focus on the multi-core architecture model proposed by Rodrigo et al. in \cite{rodrigo_2020_exploring}. This model suggests using a maximally entangled state, often called an EPR pair, as a foundational communications resource to facilitate the transmission of quantum states between cores. The communication protocol employed in this model, known as quantum teleportation, provides an efficient means to transfer quantum information between cores while leveraging classical communications.

\subsection{FGP-rOEE}
This section reviews the inter-core mapping algorithm proposed in \cite{baker_2020_timesliced}, called Fine Grained Partitioning (FGP-rOEE), which is one of the scarce algorithms proposed so far to solve the mapping problem for multi-core quantum computers. Other mapping algorithms, such as \cite{ferrari_2021_compiler} and \cite{wu_2022_autocomm}, use EPR pairs for remote operations apart from teleportation.

The FGP-rOEE algorithm takes as inputs a quantum circuit with $q$ qubits and an architecture with $N$ cores, each one containing $\frac{q}{N}$ qubits. Two assumptions are made: i) The coupling map of each core is all-to-all; i.e., all qubits in a core have direct connections to one another, allowing the direct execution of a two-qubit gate if both qubits are located within the same core. ii) Cores are interconnected all-to-all, enabling the exchange of quantum states between any pair of cores.

The algorithm's goal is to obtain a sequence of assignments of qubits to cores, one assignment per timeslice (the set of quantum gates that can be executed in parallel). The circuit is separated into timeslices, and for each of them, a valid assignment of qubits to cores is found. A valid assignment must have every pair of interacting qubits in the timeslice assigned to the same core, so no two-qubit gate involves qubits located in different cores.

For a given timeslice $t$, FGP-rOEE computes the interaction graph of that timeslice, a graph with the qubits as nodes and an edge between two nodes if the qubits interact with each other in future timeslices. The edges are weighted, representing the immediacy of the interaction. These weights are called look-ahead weights and are computed using Equation (\ref{look-ahead weights}), where $I(m, q_i,q_j) = 1$ if qubits $q_i$ and $q_j$ interact at timeslice $m$, and the exponential decay function is used so nearby timeslices have more impact on the look-ahead weights than latter ones.
\begin{equation}
    w_t(q_i, q_j) = \sum_{t < m \leq T} I(m, q_i,q_j) \cdot 2^{-(m-t)}
    \label{look-ahead weights}
\end{equation}
For qubits that interact exactly at timeslice $t$, a weight of infinity is set to the edges since, at that timeslice, they must be in the same core.

A $k$-partitioning algorithm is then applied to the interaction graph, partitioning the graph into $k$ disjoint subsets of nodes. For our case, $k$ equals the number of cores $N$, and it is a requirement that all partitions have the exact same size, as cores have a fixed size, and we can only assign $\frac{q}{N}$ qubits into each core. Because of these requirements, the number of available $k$-partitioning algorithms is scarce. 

In \cite{baker_2020_timesliced}, the use of the Overall Extreme Exchange (OEE) algorithm \cite{park_1995_algorithms} is proposed. The OEE algorithm builds upon the Kernighan–Lin (KL) algorithm and expands its capabilities. In OEE, the process begins with an initial configuration of qubits, and a pair of qubits is selected for sequential exchange. This selection is based on optimizing the minimized exchange cost for the current exchange. If the exchange cost remains positive, further exchanges are performed in an attempt to improve the configuration. The algorithm terminates if the exchange cost becomes zero or negative, indicating no further improvement can be achieved. In \cite{baker_2020_timesliced}, a variation of the OEE algorithm is proposed, the rOEE or relaxed Overall Extreme Exchange. rOEE also starts with an assignment and performs exchanges of nodes until a valid partition is reached (i.e. all interacting qubits are in the same partition).

This procedure is repeated for every pair of timeslices, using the previous assignment of qubits to cores as input for the rOEE algorithm. With this, a path of valid assignments is found over the whole circuit.

\section{Hungarian Qubit Assignment}

In this section, we propose an alternative inter-core mapper to FGP-rOEE. Similar to FGP-rOEE, the Hungarian Qubit Assignment (HQA) describes how to assign qubits to cores in between timeslices and is generalized to map the whole algorithm. However, instead of distributing qubits into cores, this algorithm distributes two-qubit operations into cores, a much easier task than the former. As each two-qubit operation within a timeslice involves two distinct qubits, the algorithm's output will be a valid assignment for that particular timeslice.

\subsection{General Overview}

We start with a valid assignment of qubits to cores for timeslice $t$. The following timeslice ($t+1$) has a set of unfeasible two-qubit gates involving qubits that are currently located in different cores. This is depicted in Figure \ref{fig: assignment example} (left), where we can see five unfeasible two-qubit operations for an example architecture of four cores.

The qubits involved in the unfeasible two-qubit operations are then removed from the assignment and placed in an auxiliary vector of unassigned qubits. Now, the task is to assign each operation to a core with enough space to take in the qubits. For this assignment, we will construct a cost matrix using the cost function in Equation (\ref{cost funcion}), which assigns a cost $\mathcal{C}_t$ to each pair of unfeasible two-qubit operation $op_i$ and core $c_j$ based on how many non-local communications are needed to place both qubits involved in the operation $q_A, q_B$ into the destination core $c_j$ (i.e. one if a qubit involved in the two-qubit gate is already placed in the core, two otherwise).
\begin{equation}
    \mathcal{C}_t (op_i, c_j) = 
\begin{cases}
    \infty & \text{if } c_j \text{ is full}\\
    1 & \text{if } q_A \in c_j \text{ or } q_B \in c_j\\
    2 & \text{otherwise}
\end{cases}
\label{cost funcion}
\end{equation}
The cost matrix is used to assign a single operation to each core. To do so, we use the Hungarian algorithm \cite{kuhn_2005_the}, a highly efficient linear assignment algorithm with polynomial time complexity ($\mathcal{O} (n^3)$). Its versatility, simple implementation, and robustness make it a favoured choice in various fields, especially when a quadratic algorithm is impractical.

The goal of the assignment problem is to find the best way to assign a set of tasks (cores) to a group of resources (unfeasible two-qubit operations) while minimizing the total cost. By doing this, at each iteration, only one operation will be assigned to each core, ensuring the core's size is not exceeded.

When an operation is assigned to a core, both qubits involved are placed in the free spaces of the core. Therefore, the number of free spaces in the core decreases by two every time an operation is assigned to it. Placing both qubits in each operation into the same core ensures all two-qubit gates will be feasible in the following timeslice.

When having more operations than cores, the Hungarian algorithm only assigns one operation per core, so some operations will remain unassigned. A new cost matrix will be computed for those unassigned operations, considering the new free spaces of each core, setting a weight of infinity for those cores already full. By doing this, we ensure that no core exceeds its capacity and that the resulting assignment will be valid. Iteratively, the cost matrix is computed, and one two-qubit operation is assigned to each core with free space until all qubits are assigned to a core.

Each core must contain an even number of qubits interacting in unfeasible two-qubit gates for this approach to work. Otherwise, when assigning operations into cores, there will be a pair of qubits left to assign and two cores with exactly one free space each, making it impossible to assign an operation to the core. An auxiliary two-qubit gate involving two non-interacting qubits from the cores with an odd number of qubits involved in unfeasible operations is created to solve this, ensuring that all cores contain an even number of free spaces and that all two-qubit operations will be allocated.

\subsection{Considering future qubit interactions}
Future interactions of qubits can be added to the cost matrix to perform an assignment that further reduces the number of non-local communications. To this end, we quantify how much qubits interact in future timeslices using the same approach as in \cite{baker_2020_timesliced}, described in Equation (\ref{look-ahead weights}). In particular, the attraction force of a qubit $q_i$ to a core $c_j$ is computed as
\begin{equation}
    \texttt{attr}^q_t (q_i, c_j) = \sum_{i'=0}^q J_t(q_{i'}, c_j) \cdot w_t(q_i, q_{i'})
\label{attraction weights}
\end{equation}

where $J_t(q_{i'}, c_j) = 1$ if qubit $q_{i'}$ is in core $c_j$ at timeslice $t$, and $w_t(q_i, q_{i'})$ is computed using Eq. (\ref{look-ahead weights}).

The new cost matrix is then computed using the costs given in Equation (\ref{cost function with attraction}), where the number of non-local communications needed and the attraction forces are combined. Note that, as each operation $op_i$ involves two qubits ($q_A$ and $q_B$), the operation's attraction force to a core is the average attraction force of the involved qubits.
\begin{equation}
    \texttt{attr}^{op}_t(op_i, c_j) = \frac{\texttt{attr}^q_t(q_A, c_j) + \texttt{attr}^q_t(q_B, c_j)}{2}
\end{equation}
\begin{equation}
    \mathcal{C}_t (op_i, c_j) = 
\begin{cases}
    \infty & \text{if } c_j \text{ is full}\\
    1 - \texttt{attr}^{op}_t(op_i, c_j) & \text{if } q_A \in c_j \text{ or } q_B \in c_j\\
    2 - \texttt{attr}^{op}_t(op_i, c_j) & \text{otherwise}
\end{cases}
\label{cost function with attraction}
\end{equation}
The proposed approach cost function is completely tunable, allowing for more complex variations of the problem. For example, if not all cores are connected to each other, the number of non-local communications to move a qubit to a given core may be more than one. The cost function can be adapted to this case, and many others, leading to a robust and widely applicable inter-core mapping algorithm.

\section{Results and Discussion}
In this section, we compare the state-of-the-art inter-core mapper (FGP-rOEE) and the proposed approach (HQA), using the number of inter-core communications as the evaluation metric (the lower, the better). For this evaluation, we will partition various quantum algorithms into modular architectures, varying the number of qubits and cores. The selected benchmarks encompass a wide variety of quantum algorithms, including Quantum Random Circuit (with varying two-qubit gate densities), GHZ State Preparation, Cuccaro Adder, Quantum Fourier Transform, Quantum Volume, and Grover's Algorithm. This diverse set of benchmarks ensures a comprehensive assessment of the mapping algorithms' performance across different quantum computation scenarios.

\begin{figure}[h]
\centering
\includegraphics[width=0.45\textwidth]{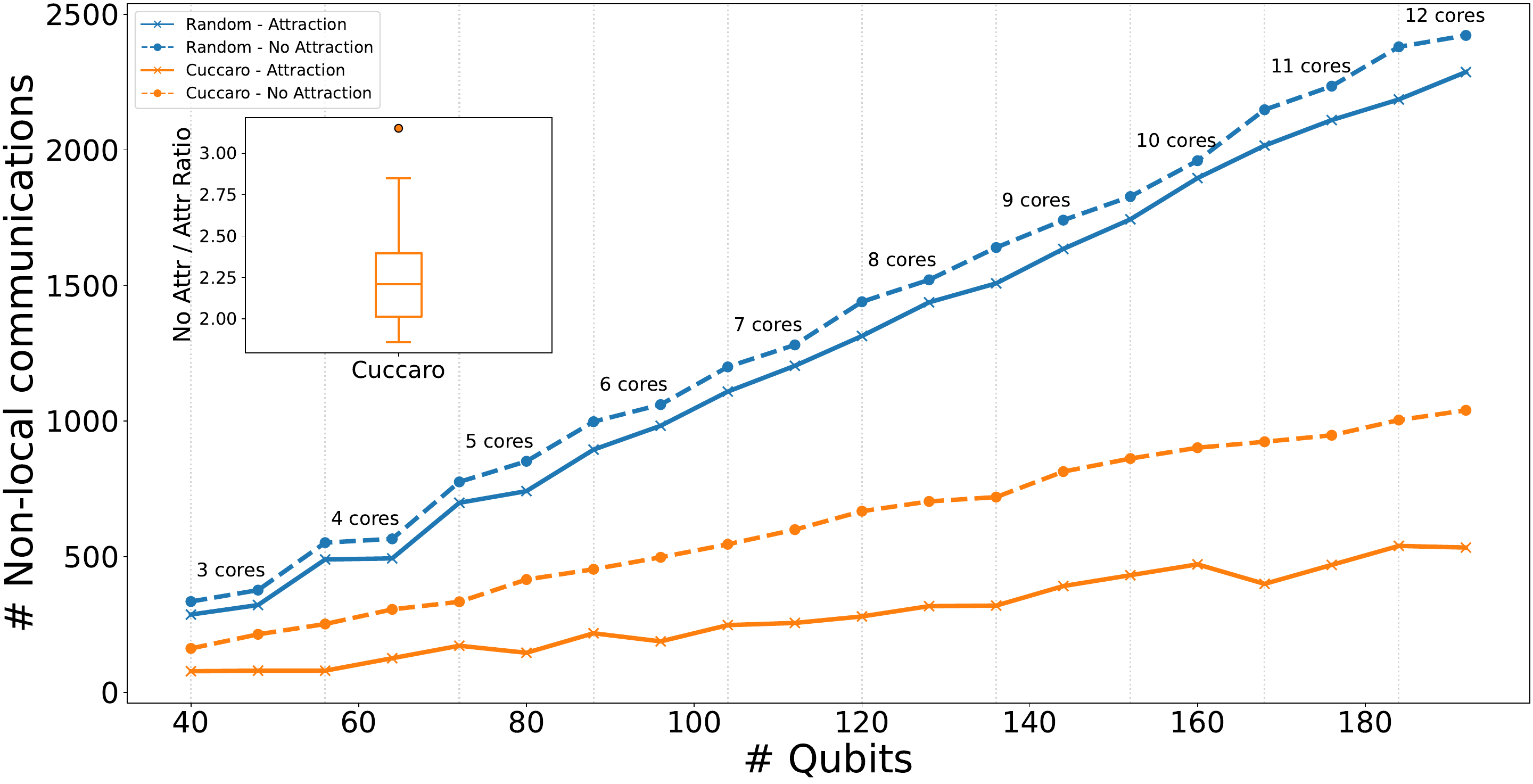}
\vspace{-0.2cm}
\caption{HQA communications with and without using the attraction force, mapped into an architecture with 16 qubits per core and as many cores as needed, depending on the number of qubits used ($x$-axis). A structured circuit (Cuccaro Adder) and an unstructured one (Random) have been selected to highlight the importance of the attraction force. The box plot depicts the Cuccaro Adder communications ratio with the two approaches.}
\label{Img: Comparison results - attraction force}
\vspace{-0.2cm}
\end{figure}

Algorithms are implemented using OpenQL \cite{openql}, as in \cite{rodrigo_2020_exploring}. They are divided into time slices and partitioned using both state-of-the-art and our proposed method. The number of non-local communications is calculated, and the ratio of our approach to the state-of-the-art indicates optimization (ratio $> 1$) or diminished performance (ratio $< 1$). For the HQA, the cost function considering future qubit interactions (defined in Equation (\ref{cost function with attraction})) is selected, as it achieves a lower number of communications than its counterpart (Equation (\ref{cost funcion})). Such comparison is depicted in Figure \ref{Img: Comparison results - attraction force}.

The performance of the mapping algorithms is assessed through two scalability approaches: one involves increasing the number of cores in a fixed-size architecture, while the other entails expanding the number of qubits in an architecture with a fixed number of cores.

\begin{figure}[h]
\centering
\includegraphics[width=0.45\textwidth]{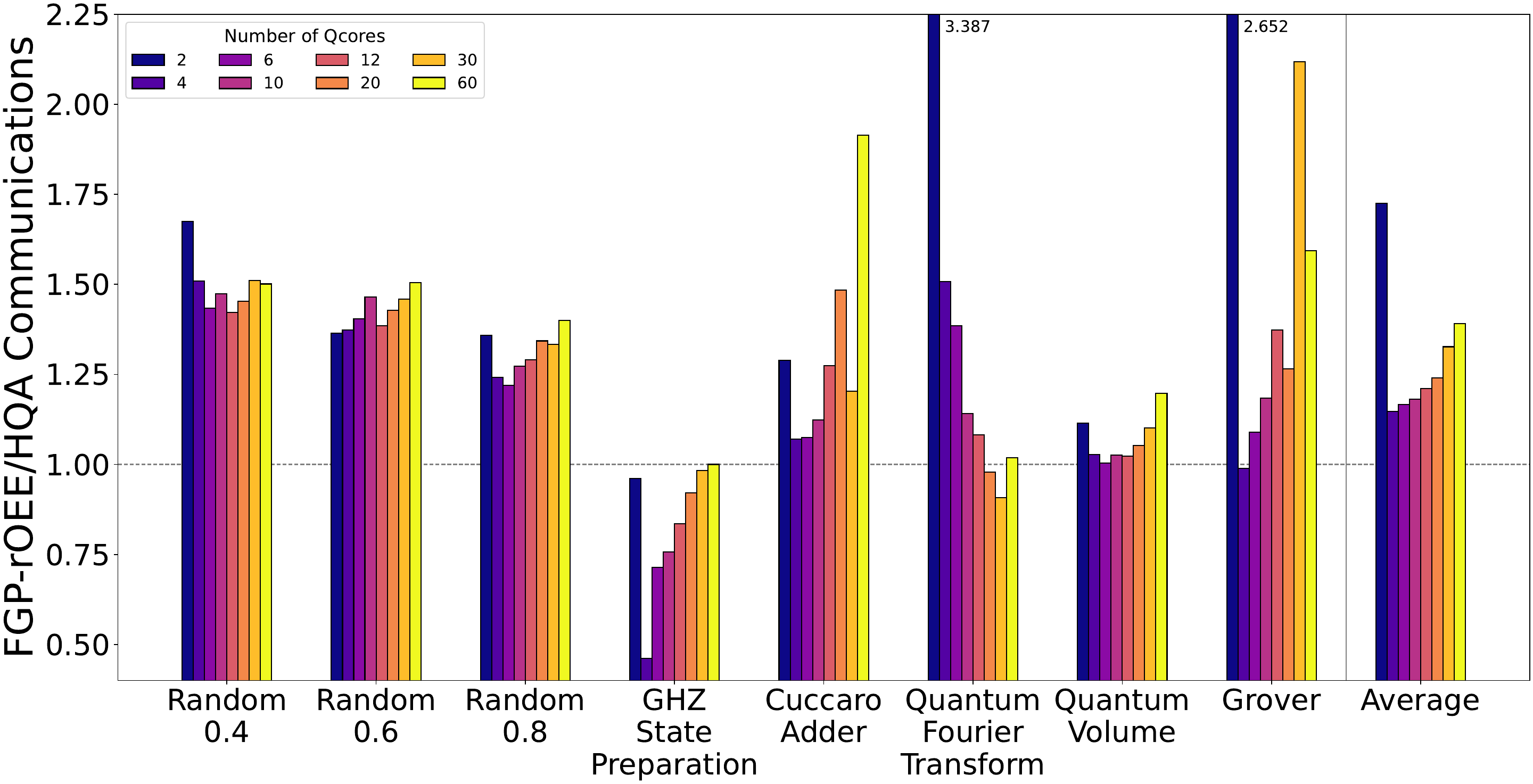}
\vspace{-0.2cm}
\caption{FGP-rOEE and HQA non-local communications ratio for the selected benchmarks, when mapping the 120-qubits into different modular architectures. Each bar represents the number of cores in the architecture, and it is selected such that each core contains the same number of qubits and ensures an even number of qubits per core. This criterion is essential to ensure the feasibility of assigning two-qubit gates within the chosen core configuration.}
\vspace{-0.2cm}
\label{Img: Comparison results - ratio}
\end{figure}

In Figure \ref{Img: Comparison results - ratio}, we show the ratio of non-local communications needed when having a fixed-size circuit (120 qubits) and mapping it into a modular architecture with a varying number of cores. Note that the number of qubits per core is also conditioned on the number of cores of the architecture since the total number of qubits matches the circuit's size.

Despite, for some cases, having a higher number of non-local communications (ratio $< 1.0$), HQA exhibits superior performance compared to FGP-rOEE, achieving improvements up to $\sim 3.4 \times$ and an overall average improvement of $\sim 1.3 \times$.

\begin{figure}[h]
\centering
\includegraphics[width=0.45\textwidth]{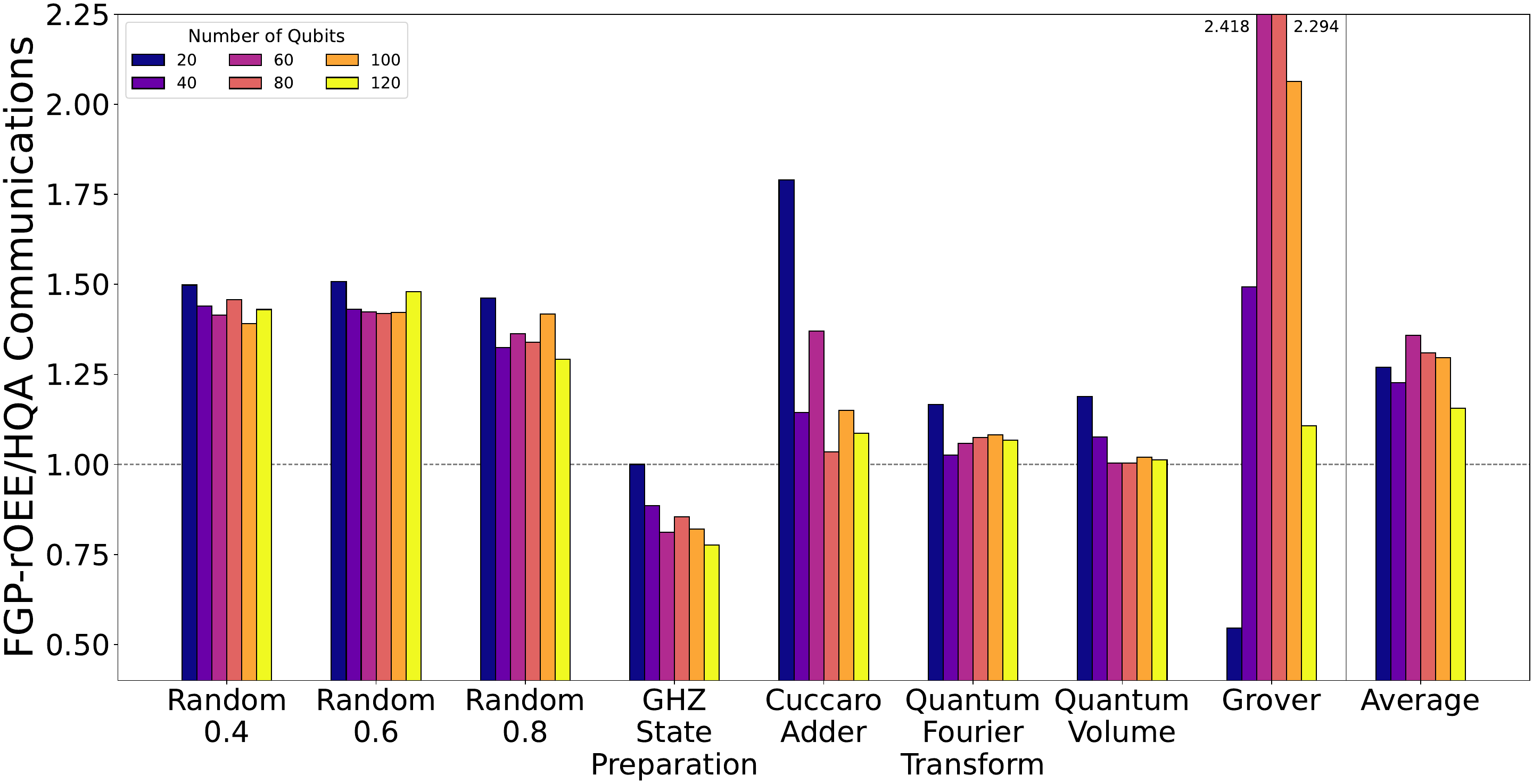}
\vspace{-0.2cm}
\caption{FGP-rOEE and HQA non-local communications ratio for the selected benchmarks, mapped into a 10-core architecture, when increasing the total number of qubits in the architecture.}
\label{Img: Comparison results - strong scaling}
\vspace{-0.2cm}
\end{figure}

Figure \ref{Img: Comparison results - strong scaling} shows the number of non-local communications needed to map the selected benchmarking algorithms into a 10-core architecture when increasing the total number of qubits and, therefore, the number of qubits per core.

It can be seen that, for both scaling approaches, the proposed algorithm usually outperforms the state-of-the-art mapping algorithm, obtaining a lower number of non-local communications, an essential aspect of multi-core quantum compilation. Unlike most benchmarks, the HQA performs worse than FGP-rOEE when mapping the GHZ circuit. We estimate this is due to the edge-case structure the GHZ circuit has, where one single qubit interacts with all other qubits sequentially. In such cases, a mapping algorithm that uses a partitioning approach instead of an assignment approach performs better.

It is worth mentioning that the attraction force for HQA has been designed to match the information used in FGP-rOEE by using the same look-ahead weights. However, several look-ahead functions are evaluated in \cite{baker_2020_timesliced}, and the best-performing one is selected. Therefore, tuning the attraction force of the proposed approach has the potential to decrease the number of non-local communications even further.

\section{Conclusions}
Modular approaches are envisioned to appear very soon in quantum computers, bringing new challenges in compiling quantum algorithms to be adapted to modular architectures. In this work, we have proposed an algorithm to distribute, along the circuit, each qubit to a partition so each two-qubit gate can be executed. The proposed algorithm produces assignments based on simplified premises than other modular-architecture mappers, as based on the assignments of unfeasible two-qubit gates into cores, instead of partitioning the interaction graph, as it was done by \cite{baker_2020_timesliced}. By doing this, a lower number of non-local communications between cores is achieved, improving the mapping of the quantum algorithm. Moreover, the proposed algorithm is highly customizable, allowing for a different topology among cores (not only all-to-all connectivity). A different definition of the attraction force of qubits to cores can be proposed, for this work, we have used the same information that FGP-rOEE use. 

\end{document}